\documentclass[aps,prx,reprint,superscriptaddress,nobibnotes,amsmath,amssymb,floatfix,showkeys,raggedbottom]{revtex4-2}

\usepackage{graphicx}
\usepackage{amssymb}
\usepackage{mathtools}
\DeclareMathOperator{\asinh}{asinh}
\usepackage{microtype}
\usepackage{booktabs}
\usepackage{hyperref}
\usepackage{url}
\hypersetup{
	colorlinks=true,
	linkcolor=blue,
	citecolor=blue,
	urlcolor=blue,
	breaklinks=true,
	pdflang={en-US},
	pdftitle={Laser-Polarization Dependence of Betatron X-Ray Production and Angular Collection in Laser-Wakefield Acceleration},
	pdfauthor={ A. C. Berceanu, V. Horny, D. Stutman, and P. Tomassini },
	pdfsubject={Particle-in-cell study of polarization-dependent electron dynamics and betatron x-ray emission},
	pdfkeywords={laser-wakefield acceleration, VHEE, betatron radiation, laser-plasma accelerator, particle-in-cell simulation}
}
\begin{document}
\title{Laser-Polarization Dependence of Betatron X-Ray Production and Angular Collection in Laser-Wakefield Acceleration}
\author{A. C. Berceanu}
\email{andrei.berceanu@eli-np.ro}
\affiliation{Extreme Light Infrastructure -- Nuclear Physics (ELI-NP) and Horia Hulubei National Institute for R\&D in Physics and Nuclear Engineering (IFIN-HH), 30 Reactorului Street, 077125 M\u{a}gurele, Romania}
\author{V. Horn{\'y}}
\affiliation{Extreme Light Infrastructure -- Nuclear Physics (ELI-NP) and Horia Hulubei National Institute for R\&D in Physics and Nuclear Engineering (IFIN-HH), 30 Reactorului Street, 077125 M\u{a}gurele, Romania}
\author{D. Stutman}
\affiliation{Extreme Light Infrastructure -- Nuclear Physics (ELI-NP) and Horia Hulubei National Institute for R\&D in Physics and Nuclear Engineering (IFIN-HH), 30 Reactorului Street, 077125 M\u{a}gurele, Romania}
\author{P. Tomassini}
\email{paolo.tomassini@eli-np.ro}
\affiliation{Extreme Light Infrastructure -- Nuclear Physics (ELI-NP) and Horia Hulubei National Institute for R\&D in Physics and Nuclear Engineering (IFIN-HH), 30 Reactorului Street, 077125 M\u{a}gurele, Romania}
\affiliation{ELI Beamlines Facility, Extreme Light Infrastructure ERIC, Za Radnici 835, 25241 Doln{\'\i} B\v{r}e\v{z}any, Czech Republic}

\begin{abstract}
	\mdseries\unboldmath
	Laser polarization affects both the production of betatron x rays and their angular distribution, which together determine the photon yield collected within a fixed angular acceptance.
	We investigate these effects using particle-in-cell simulations of laser-wakefield acceleration with linearly polarized (LP) and circularly polarized (CP) pulses of equal energy in helium--nitrogen targets of two different lengths.
	In the longer target, CP produces more photons early in propagation, whereas LP produces more later.
	For 20--60-keV photons transmitted through 2~mm of aluminum, the LP/CP yield ratio before angular selection changes from 0.88 early on to 1.91 in a late stage.
	Late-emitting LP electrons develop pronounced transverse oscillations along the laser polarization direction, accompanied by angular broadening of the radiation.
	This broadening largely removes the late LP production advantage within a central $\pm5$-mrad acceptance, where the two yields are nearly equal.
	Integrated over the full interaction, however, LP supplies more central photons, while CP provides more uniform illumination.
	In the shorter target, the extracted charges above 100~MeV are nearly equal, yet transport through a quadrupole triplet delivers 750~pC for LP and 500~pC for CP in the 350--500-MeV energy band within a radius of 5~mm.
	Collected photon and electron yields are thus shaped by propagation length and collection acceptance.
\end{abstract}
\maketitle

\section{Introduction}
\label{intro}

Laser-wakefield acceleration (LWFA)~\cite{TajimaDawson1979} produces relativistic electron bunches that emit betatron x rays as they oscillate in the focusing fields of the plasma wake~\cite{Rousse2004}.
The radiation spectrum depends on electron energy and trajectory curvature, while its angular distribution reflects the electron motion~\cite{Esarey2002,Corde2013}.
Wood et al.~\cite{Wood2026} demonstrated enhanced x-ray emission beyond laser depletion, dominated by an electron bunch injected later that has lower energy and a larger betatron amplitude.
This emission after laser depletion illustrates why interpreting the x-ray output requires following electron energy and transverse motion throughout the propagation.

Control of transverse motion through electron steering~\cite{Yu2018} or staged acceleration--radiation schemes~\cite{Ferri2018} can enhance betatron emission, while injection conditions~\cite{Horny2017} and modulation of the oscillations by the laser~\cite{Horny2020} shape the x-ray pulse in time.
Ionization-injection models relate laser polarization to the residual momentum, emittance, and trapping of the injected electrons~\cite{Schroeder2014,Tomassini2022,Chen2012}.
Vieira et al.~\cite{Vieira2016} demonstrate control of x-ray polarization through helical electron motion driven by circularly or elliptically polarized lasers.
Photon production and angular collection provide a complementary comparison of linearly polarized (LP) and circularly polarized (CP) drivers, distinct from the polarization state of the emitted x rays.

Source comparisons must distinguish total photon production from the photon number accepted by an imaging system and separately quantify the electron charge delivered by a transport line.
Spatial coherence enables phase contrast with polychromatic x rays~\cite{Wilkins1996}, as illustrated by betatron imaging of biological samples~\cite{Fourmaux2011,Kneip2011,Chaulagain2017} and quantitative tomography~\cite{Wenz2015}.
Laser-based interferometric x-ray imaging (LIXI) for mammography~\cite{Safca2022,Stutman2023} motivates our comparison of accepted photon number, illumination uniformity, and transverse coherence.
Studies of irradiation with very high-energy electrons (VHEE) connect source and transport properties to dose calculations~\cite{Glinec2006,Fuchs2009,Kalvala2026} and experimental dose control~\cite{Labate2020}.

In this work, we investigate how the relative betatron x-ray yields of LP and CP drivers of equal energy evolve during propagation and change with angular collection.
We address these questions with particle-in-cell (PIC) simulations of helium--nitrogen targets that differ only in the length of their density plateau, referred to below as the short and long targets.
In the long target, we resolve photon production in propagation distance, energy, and angle, and relate the late emission to electron trajectories and azimuthally decomposed fields.
As a complementary test of electron delivery, we transport bunches extracted from the short target through a quadrupole triplet.
The simulations reveal a reversal in the relative LP and CP photon yields during propagation, whose effect on the accumulated output depends on angular acceptance.

\begin{figure*}[!htbp]
	\centering
	\includegraphics[width=.80\textwidth]{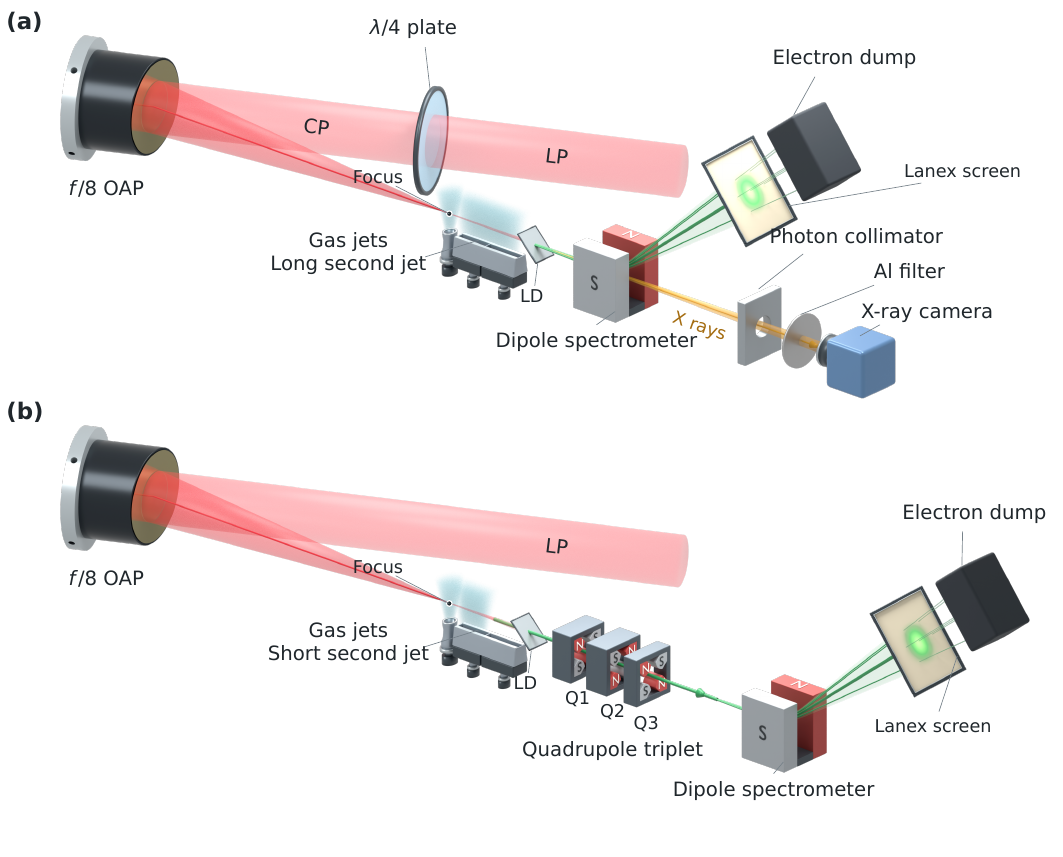}
	\caption{Conceptual layouts for (a) betatron x rays from a long target with circular laser polarization (CP) and (b) very high-energy electrons (VHEE) from a short target with linear polarization (LP).
	The $\lambda/4$ plate converts LP to CP.
	OAP denotes the off-axis parabola, LD the laser dump, Q1--Q3 the quadrupoles, and N/S the magnetic poles.
	Red, green, and amber rays represent laser light, electrons, and x rays, respectively.
	Cyan envelopes indicate the gas jets, and dots mark the position of the laser focal plane.}
	\label{fig:setup}
\end{figure*}

\section{Physical model and numerical methods}
\label{Setup}

\subsection{Laser and plasma configurations}
\label{sec:laser-target}

Figure~\ref{fig:setup} illustrates the two output configurations.
The initial design uses the scaling laws for a matched spot in the nonlinear blowout regime~\cite{Lu2007}, ultrarelativistic wakefield scaling~\cite{GordienkoPukhov2005}, and operation at high charge in the regime dominated by pump depletion~\cite{Papp2021,Horny2024}.
The laser propagates along $z$, with $x$ the direction of the LP electric field.
The model uses a pulse of 2.1~J at 810~nm with an intensity full width at half maximum (FWHM) duration of 25~fs.
An $f/8$ off-axis parabola focuses the pulse to an intensity FWHM diameter of 13~$\mu$m.
We hold both energy and duration fixed to distinguish polarization effects from changes in acceleration efficiency associated with pulse compression~\cite{Vais2024}.
A flattened Gaussian profile~\cite{Santarsiero1997} provides the same intensity envelope for both polarizations, and we introduce the ellipticity $\chi$ as the ratio of the minor to major semiaxes of the electric field ellipse.

The target has two density plateaus joined by an upramp that regulates electron dephasing~\cite{PukhovKostyukov2008,Aniculaesei2019}.
Only the length of the second plateau changes between the short and long configurations, giving gas exits at $z=2.46$~mm for the short target and $z=5.15$~mm for the long target.
The initial free-electron density is $n_e=2n_{\mathrm{He}}+5n_{\mathrm N}$, with a ratio of atomic number densities $n_{\mathrm N}/n_{\mathrm{He}}=0.3$ and nitrogen initially ionized to N$^{5+}$ by the laser pre-pulse.
The initial free-electron densities are $n_0=6.72\times10^{18}$~cm$^{-3}$ on the first plateau and $1.26\times10^{19}$~cm$^{-3}$ on the second.
We use the first-plateau value $n_0$ as the density normalization throughout.
The full laser and plasma parameters are detailed in Appendix~\ref{app:fbpic-parameters}.

The simulations use the quasi-cylindrical Fourier--Bessel particle-in-cell (FBPIC) code~\cite{Lehe2016}.
Appendix~\ref{app:numerics} provides the numerical implementation details.

\subsection{Electron and field observables}
\label{sec:numerical-model}
\begin{figure*}[!htbp]
	\centering
	\includegraphics[width=.92\textwidth]{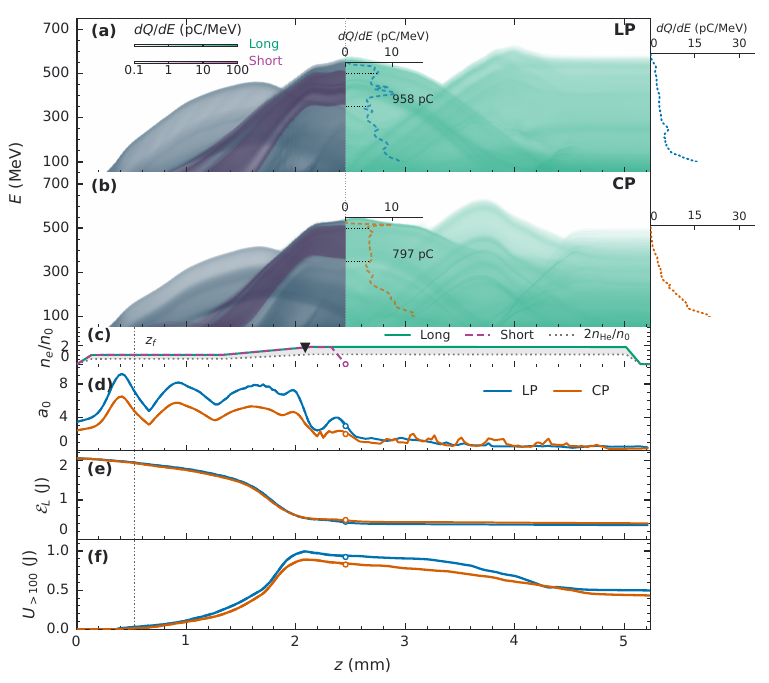}
	\caption{Laser and electron evolution in the short and long targets.
	(a) and (b) LP and CP charge spectra with $u_z>100$, versus propagation distance $z$.
	Long (green) and short (purple) maps share a logarithmic opacity scale, with the narrow bands tracing the electrons selected at the short target's exit in the 350--500-MeV band.
	Dashed insets show spectra at the gas exit and labeled band charges, and dotted profiles show spectra for the long target at $z=5.24$~mm.
	(c) Initial density $n_e/n_0$, with $n_0=6.72\times10^{18}$~cm$^{-3}$: dotted helium and shaded nitrogen contributions.
	The triangle marks the propagation position used in Fig.~\ref{fig:snapshot_betatron}.
	(d) Peak normalized laser amplitude $a_0$.
	(e) Remaining laser energy $\mathcal{E}_L$.
	(f) Instantaneous electron kinetic energy $U_{>100}$ above 100~MeV.
	Curves in (d)--(f) use LP blue, CP orange, long solid, and short dashed.
	Open circles mark the exit of the short target at 2.46~mm.
	Vertical dotted guides mark this exit in (a) and (b) and the laser focal plane at $z_f$ in (c)--(f).}
	\label{fig:Betatron_global}
\end{figure*}

Figure~\ref{fig:Betatron_global} compares the electron energy spectra and total kinetic energy with the plasma density and laser evolution for both target lengths and polarizations.
Electron kinetic energy is defined as $E=(\gamma-1)m_ec^2$, where $u_i=p_i/(m_ec)$ and $\gamma=\sqrt{1+u_x^2+u_y^2+u_z^2}$ is the Lorentz factor.
Here $m_e$, $c$, and $e>0$ denote the electron mass, speed of light, and elementary charge, respectively, while $p_i$ denotes the electron momentum.
Unless photon weights are specified, a scalar particle quantity $h$ has the mean $\langle h\rangle=\sum_jq_jh_j/\sum_jq_j$, with $q_j>0$ the charge magnitude represented by simulation particle $j$.
For quantities such as position or momentum, the centered rms width $\sigma_h=\sqrt{\langle h^2\rangle-\langle h\rangle^2}$ is the standard deviation about the mean.
Electron angles are $\theta_i=\tan^{-1}(p_i/p_z)$, $i=x,y$.
The spectra retain electrons with $u_z>100$ and use the propagation coordinate $z\simeq ct$, where $t$ is laboratory time.
Within this $u_z>100$ population, the instantaneous total kinetic energy above 100~MeV is $U_{>100}=\sum_{j:\,E_j>100\,\mathrm{MeV}}(q_j/e)E_j$, evaluated in the moving window using native bin-center energies for the long target.
Spectra at the exit of the short target count electron crossings of a fixed plane and supply the input for the beam transport system.
Plots for the short target stop at that plane, whereas those for the long target end at the final snapshot of the moving window, $z=5.24$~mm.

Figure~\ref{fig:snapshot_betatron} shows simultaneous slices at $y=0$ in the laboratory frame of the net charge density $\rho$, longitudinal electric field $E_z$, and transverse field $W_x=E_x-cB_y$.
The field comparison separates longitudinal energy exchange from transverse confinement~\cite{Esarey2002,Corde2013}.
For forward ultrarelativistic electrons, $F_x\simeq-eW_x$ when $v_z\simeq c$ and $v_yB_z$ is negligible.
The laser amplitude is $a_0=eE_{\mathrm{pk}}/(m_ec\omega_0)$, where $E_{\mathrm{pk}}$ is the peak major semiaxis of the analytic transverse field's polarization ellipse and $\omega_0=2\pi c/\lambda_0$ is the nominal angular frequency.
Both laser observables use longitudinal wave numbers $0.30\leq |k_z|/k_0\leq1.70$, where $k_0=\omega_0/c$.
The remaining laser energy $\mathcal E_L$ is the electromagnetic energy in this interval.

\subsection{Radiation model and photon observables}
\label{sec:photon-methods}
\label{sec:trajectory-methods}

We developed a radiation diagnostic for FBPIC simulations in a Lorentz-boosted frame by extending its synchrotron module for the laboratory frame~\cite{FBPICbetatron}.
It retains the local incoherent strong-wiggler model~\cite{Corde2013}, transforming emission events to the laboratory frame and accumulating photon yields by laboratory emission time.
Appendix~\ref{app:radiation-model} summarizes the implementation, with a detailed account planned for a forthcoming publication.
All photon yields and distributions are reported per laser shot unless otherwise specified, with $N_\gamma$ denoting photon number, $E_\gamma$ photon energy, and ``ph'' abbreviating photons.

The spectral density $S_\gamma=\mathrm{d}^2N_\gamma/(\mathrm{d}E_\gamma\,\mathrm{d}z)$ resolves production along propagation. We distinguish total production before angular selection from photons within the angular acceptance $|\theta_x|,|\theta_y|\leq50$~mrad and its central $\pm5$-mrad region, where the projected angles refer to the $xz$ and $yz$ planes.
Motivated by LIXI phase-contrast mammography setups, we compare photon yields in the 20--60-keV band after transmission through an aluminum filter of 2~mm thickness, which suppresses low-energy photons.
We use 35~keV as a representative energy for the spatial maps and evaluate coherence at this energy approximately 5.7~m downstream of the plasma, matching the total length of a reference Talbot--Lau interferometer~\cite{Safca2022}.
We also consider other possible filter thicknesses, and
Appendix~\ref{app:radiation-model} describes the model for transmission through aluminum and gives the native resolutions in photon energy and angle.

Spatial emission maps at 35~keV with a relative bandwidth of 0.1\% accumulate photons within the central region at their emission positions $(x,y)$.
Their density is $\overline{\Sigma}_{\rm em}=\Delta N_\gamma/(\Delta z\,\Delta x\,\Delta y)$ for a propagation interval $\Delta z$ and pixel area $\Delta x\,\Delta y$.
The accumulated angular fluence is $\Sigma_\theta=\mathrm{d}^2N_\gamma/(\mathrm{d}\theta_x\,\mathrm{d}\theta_y)$, with its coefficient of variation defined as the standard deviation divided by the mean within the selected region.
To characterize the effective source, we project rays emitted at $(x,y,z_{\rm em})$ onto a plane at $z_{\mathrm{s}}=2.7$~mm using
$X=x+(z_{\mathrm{s}}-z_{\rm em})\tan\theta_x$ and $Y=y+(z_{\mathrm{s}}-z_{\rm em})\tan\theta_y$.
Here $z_{\rm em}$ locates emission in the laboratory, whereas $z\simeq ct$ tracks the window front.
This plane lies near the positions where the projected 35-keV photon distributions have their smallest transverse rms widths for both LP and CP.
Therefore, the resulting $\Sigma_{\mathrm{s}}=\mathrm{d}^2N_\gamma/(\mathrm{d}X\,\mathrm{d}Y)$ combines emission positions and directions at that plane.

We estimate betatron source coherence using the complex vector radiation amplitude $\mathbf A_j(\mathbf R,E_\gamma)$, integrated along each electron trajectory, assuming statistically independent emission from different electrons.
Spatial coherence is calculated by correlating each electron's radiation field at two observation points and summing these correlations over the whole electron ensemble.
For observation points $\mathbf R_1$ and $\mathbf R_2$, the cross-spectral density summed over field components and its normalized degree of coherence are
\begin{equation}
\begin{aligned}
 G_{ab}(E_\gamma)&=\sum_j W_j\,\mathbf A_j(\mathbf R_a,E_\gamma)
 \mathbin{\cdot}\mathbf A_j^*(\mathbf R_b,E_\gamma),\\
 \mu_{12}(E_\gamma)&=\frac{G_{12}(E_\gamma)}{\sqrt{G_{11}(E_\gamma)G_{22}(E_\gamma)}}.
\end{aligned}
\label{eq:transverse-coherence}
\end{equation}
Here $a,b\in\{1,2\}$ and $W_j=w_j/\pi_j$, with $w_j$ the number of electrons represented by simulation particle $j$ and $\pi_j$ its trajectory sampling probability.
We use $\mu\equiv\mu_{12}$ below.
At 35~keV, we evaluate horizontal and vertical point pairs centered on the nominal propagation axis in the observation plane approximately 5.7~m downstream of the plasma.
Their transverse separation is $\Delta=|\mathbf R_2-\mathbf R_1|$, and the coherence length is its value when $|\mu_{12}|$ first falls to $e^{-1/2}$.
At this energy, the modeled aluminum filter multiplies $G_{12}$, $G_{11}$, and $G_{22}$ by the same transmission factor.

\section{Laser and electron evolution}
\label{sec:accelerator-dynamics}
\label{sec:energy-evolution}

\begin{figure}[!htbp]
	\centering
	\includegraphics[width=\columnwidth]{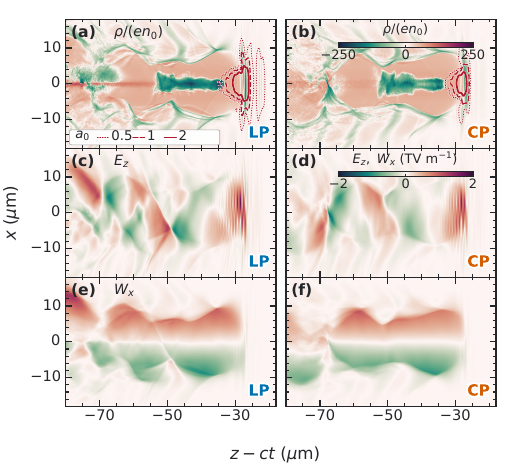}
	\caption{Plasma in the long target at $t=6.97$~ps ($z=2.09$~mm): LP left, CP right.
	(a) and (b) Net charge density $\rho/(e n_0)$, with $n_0=6.72\times10^{18}$~cm$^{-3}$.
	(c) and (d) Longitudinal electric field $E_z$.
	(e) and (f) Transverse focusing field $W_x=E_x-cB_y$.
	Red contours of the laser envelope mark $a_0=0.5,1,2$ (dotted, dashed, and solid, respectively).
	The color scales use inverse hyperbolic sine ($\asinh$) scaling for charge density and linear scaling for fields.}
	\label{fig:snapshot_betatron}
\end{figure}

The shared plasma profile gives the short and long targets nearly identical early electron evolution [Figs.~\ref{fig:Betatron_global}(a)--\ref{fig:Betatron_global}(c)].
The peak laser field initially rises through nonlinear self-focusing and pulse reshaping~\cite{Esarey1997,Vieira2010}, reaching approximately twice its nominal amplitude near $z=0.41$~mm. A narrower beam or shorter pulse can raise its peak amplitude even as the total pulse energy decreases.
As the remaining laser energy $\mathcal E_L$ falls through the density upramp, the electron spectra extend to higher energies, and the instantaneous total kinetic energy $U_{>100}$ of electrons above 100~MeV reaches its maximum near $z=2.09$~mm, where Fig.~\ref{fig:snapshot_betatron} compares the fields.
The short target extracts the bunch after this maximum, whereas the long target allows the surviving electrons to continue interacting with the evolving wake.

Near this maximum, both polarizations form a dense axial electron column inside a wider cavity depleted of electrons [Figs.~\ref{fig:snapshot_betatron}(a) and~\ref{fig:snapshot_betatron}(b)].
Negative $E_z$ accelerates electrons forward [Figs.~\ref{fig:snapshot_betatron}(c) and~\ref{fig:snapshot_betatron}(d)].
The transverse force retains a broader restoring pattern, with $W_x>0$ above the axis and $W_x<0$ below it, directing electrons inward and driving the betatron oscillations that produce the radiation [Figs.~\ref{fig:snapshot_betatron}(e) and~\ref{fig:snapshot_betatron}(f)].

At the exit of the short target, the 350--500-MeV spectral band contains a charge of 958~pC (LP) and 797~pC (CP).
The purple narrow band highlights in Figs.~\ref{fig:Betatron_global}(a) and~\ref{fig:Betatron_global}(b) trace the earlier injection and evolution of these electron bunches.
Continued interaction in the long target redistributes the spectrum as $U_{>100}$ decreases, with LP retaining more charge in this band and developing stronger horizontal broadening (Appendix~\ref{app:electron-evolution}).
In Sec.~\ref{Betatron}, we follow the radiation produced during this continued interaction.
In Sec.~\ref{VHEE}, we characterize the bunches from the short target as they propagate through the quadrupole triplet.

\section{Betatron x-ray emission during propagation}
\label{Betatron}

\subsection{Spectral, angular, and spatial evolution}
\label{sec:matched-mechanism}

\begin{figure*}[!htbp]
\centering
\includegraphics[width=.90\textwidth]{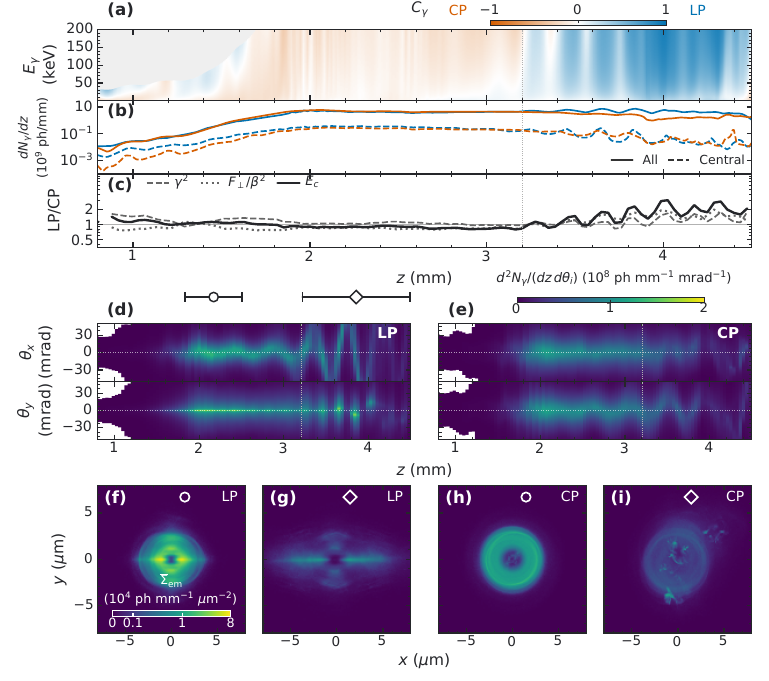}
\caption{Betatron radiation during propagation in the long target: LP blue/left, CP orange/right.
(a) Contrast in spectral density before angular selection, $C_\gamma=(S_\gamma^{\rm LP}-S_\gamma^{\rm CP})/(S_\gamma^{\rm LP}+S_\gamma^{\rm CP})$, where $S_\gamma=\mathrm{d}^2N_\gamma/(\mathrm{d}E_\gamma\,\mathrm{d}z)$ is the photon yield per unit energy and propagation distance.
(b) 20--60-keV yield per unit propagation: total (solid), central $\pm5$-mrad region (dashed).
(c) LP/CP ratios of the geometric means of $\gamma^2$, $F_\perp/\beta^2$, and $E_c$, weighted by photon yield.
(d) and (e) LP and CP 20--60-keV photon angular distributions, respectively, with $\theta_x$ above and $\theta_y$ below.
Each distribution is integrated over the other projected angle within $\pm50$~mrad.
(f)--(i) Early (circle) and late (diamond) spatial emission densities $\overline{\Sigma}_{\rm em}$ at 35~keV with a relative bandwidth of 0.1\% within the central region, with shared $\asinh$ scaling and intervals shown above panel (d).
Panels (b)--(i) use an aluminum filter of 2~mm thickness.}
\label{fig:tracks}
\label{fig:matched-mechanism}
\end{figure*}

The relative LP and CP photon yields reverse during propagation: CP emits more strongly early on, whereas late LP emission is stronger and more energetic [Figs.~\ref{fig:tracks}(a) and~\ref{fig:tracks}(b)].
After transmission through 2~mm of aluminum and before angular selection, the 20--60-keV LP/CP yield ratio rises from 0.88 over $z=1.85$--$2.50$~mm to 1.91 over $z=3.20$--$4.50$~mm.
This reversal trend is even stronger at higher photon energy, indicating spectral hardening as well as greater photon production.

The synchrotron critical energy connects this hardening to both electron acceleration and increased trajectory curvature~\cite{Esarey2002,Corde2013}.
For the implemented radiation model,
\begin{equation}
 E_c=\frac{3\hbar}{2m_ec}\,\frac{\gamma^2F_\perp}{\beta^2}.
\label{eq:critical-energy}
\end{equation}
Here $\beta=|\mathbf v|/c$ and
$F_\perp=|\mathbf F-(\mathbf F\cdot\hat{\mathbf v})\hat{\mathbf v}|$ is the force normal to the trajectory, with
$\mathbf F=-e(\mathbf E+\mathbf v\times\mathbf B)$ the Lorentz force, and $\hat{\mathbf v}=\mathbf v/|\mathbf v|$.
At fixed electron energy, stronger bending raises the critical photon energy.
To separate these two factors, we weight each electron by its contribution to the 20--60-keV photon yield in the corresponding bin of emission time.
For a positive electron quantity $h$, we denote the mean with these weights by $\langle h\rangle_\gamma$ and the geometric mean by $\mathcal M_\gamma(h)=\exp[\langle\ln h\rangle_\gamma]$.
Using identical normalized photon weights for all three quantities within each polarization gives the exact identity
\begin{equation}
 \mathcal M_\gamma(E_c)=\frac{3\hbar}{2m_ec}\,\mathcal M_\gamma(\gamma^2)\,\mathcal M_\gamma(F_\perp/\beta^2).
\label{eq:energy-bending}
\end{equation}
This identity also holds when electron energy and force are correlated.
Figure~\ref{fig:tracks}(c) compares the resulting LP/CP ratios of the geometric means: the ratio for $E_c$ equals the product of the ratios for $\gamma^2$ and $F_\perp/\beta^2$.

Early CP emitters experience stronger normal forces, outweighing their slightly lower energies [Fig.~\ref{fig:tracks}(c)].
Later LP emitters combine higher energy with stronger bending, and the LP/CP ratio of critical energies reaches 1.50.
The higher critical energy of the later LP emission thus reflects both electron energy and transverse bending.

Angular selection favors LP over CP early in propagation and removes most of the later LP production advantage.
Over $z=1.85$--$2.50$~mm, the 20--60-keV LP/CP yield ratio (after transmission through aluminum) is 0.88 before angular selection but 1.35 within the central $\pm5$-mrad region.
Over $z=3.20$--$4.50$~mm, the ratio before angular selection rises to 1.91, whereas the central yields are approximately equal [Fig.~\ref{fig:tracks}(b)].

Relativistic radiation is directed in a forward cone around the instantaneous electron velocity~\cite{Corde2013}, so increased transverse motion can strengthen emission while moving it away from the axis.
The LP enhancement toward the end of the interaction is accompanied by stronger horizontal angular broadening [Figs.~\ref{fig:tracks}(d) and~\ref{fig:tracks}(e)].
Over $z=3.20$--$4.50$~mm, the $\pm50$-mrad angular acceptance contains 61\% of the filtered 20--60 keV LP photons, compared with approximately 82\% for CP.

At 35~keV within the central square region, CP emission evolves from an annular distribution toward a less hollow one, whereas LP emission broadens mainly in $x$ [Figs.~\ref{fig:tracks}(f)--\ref{fig:tracks}(i)].
Therefore, these maps locate the emission inside the plasma: the angular redistribution is accompanied by a change in the shape of the emitting region, with LP widening preferentially along the laser polarization direction.

\subsection{Transverse dynamics of late-emitting electrons}
\label{sec:angular-acceptance}

\begin{figure}[!htbp]
\centering
\includegraphics[width=\columnwidth]{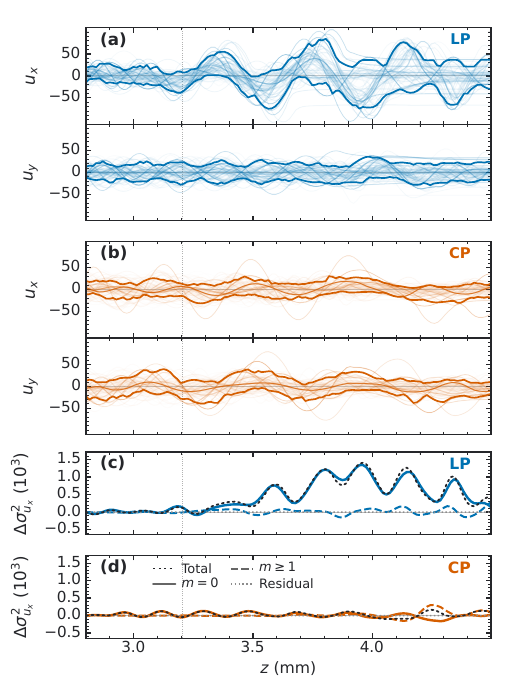}
\caption{Dynamics of electrons contributing to the late betatron emission in the long target: LP in (a) and (c), CP in (b) and (d).
Each electron has a fixed weight given by its 20--60 keV photon yield within $\pm50$~mrad after transmission through 2~mm of aluminum over $z=3.20$--$4.50$~mm (dotted vertical lines).
(a) and (b) Transverse momentum histories, with $u_x$ above and $u_y$ below. Opacity represents photon weights, and bounds mark the 10th and 90th percentiles.
(c) and (d) Contributions of azimuthal modes to the horizontal variance change $\Delta\sigma_{u_x}^2$ from $z_0=2.8$~mm.}
\label{fig:late-emitter-dynamics}
\end{figure}

To identify the motion accompanying the stronger LP emission and angular broadening, we follow a fixed group of electrons contributing to the 20--60 keV, $\pm50$-mrad selection.
Each sampled electron is assigned a fixed weight equal to its contribution to the filtered photon yield over $z=3.20$--$4.50$~mm and its trajectory is followed from $z=2.8$~mm.
The LP electrons develop pronounced horizontal momentum oscillations, whereas the CP electrons exhibit more balanced motion in the two transverse directions [Figs.~\ref{fig:late-emitter-dynamics}(a) and~\ref{fig:late-emitter-dynamics}(b)].
At $z=3.96$~mm, the ratio of centered momentum widths $\sigma_{u_x}/\sigma_{u_y}$ reaches 2.43 for LP and 1.06 for CP.
Holding the population and weights fixed connects this anisotropy to the evolving motion of those electrons.

Decomposing the fields into azimuthal modes~\cite{Lehe2016,Miller2023,BerceanuDelDotto2026} connects the fields sampled along these trajectories to their changing transverse momenta.
Integrating the force from azimuthal mode $m$ along each trajectory gives its contribution to the normalized momentum change,
\[
 \Delta u_{m,x}=\frac{1}{m_ec}\int_{t_0}^{t}F_{m,x}\,\mathrm{d}t,
\]
with initial momentum $u_{x,0}$ at $z_0=2.8$~mm.
Here the averaging convention in Sec.~\ref{sec:numerical-model} uses the fixed photon weights $w_j^{\mathrm{ph}}$ in place of charge weights, giving $\langle h\rangle_{\mathrm{ph}}=\sum_jw_j^{\mathrm{ph}}h_j/\sum_jw_j^{\mathrm{ph}}$ and $\delta h=h-\langle h\rangle_{\mathrm{ph}}$.
The corresponding momentum variance is $\sigma_{u_x}^2=\langle(\delta u_x)^2\rangle_{\mathrm{ph}}$.
We denote the contribution of each mode to its change by
\begin{equation}
\begin{aligned}
 \mathcal V_{m,x}={}&2\langle\delta u_{x,0}\,\delta(\Delta u_{m,x})\rangle_{\mathrm{ph}}\\
       &+\langle\delta(\Delta u_x)\,\delta(\Delta u_{m,x})\rangle_{\mathrm{ph}}.
\end{aligned}
\label{eq:modal-variance}
\end{equation}
Here $\Delta u_x=u_x(t)-u_{x,0}=\sum_m\Delta u_{m,x}+\Delta u_{\mathrm{res},x}$.
The residual $\Delta u_{\mathrm{res},x}$ is the difference between the actual momentum change and the summed mode integrals, arising from field reconstruction and temporal integration.
Replacing $\Delta u_{m,x}$ by $\Delta u_{\mathrm{res},x}$ in Eq.~\eqref{eq:modal-variance} defines its contribution $\mathcal V_{\mathrm{res},x}$.
Together, they sum to the variance change $\Delta\sigma_{u_x}^2=\sigma_{u_x}^2(t)-\sigma_{u_x}^2(t_0)$, sharing cross terms symmetrically.

Positive and negative terms represent broadening and narrowing contributions, respectively, while retaining correlations among modes.
For this fixed cohort, the axisymmetric ($m=0$) contribution defined in Eq.~\eqref{eq:modal-variance} accounts for most of the increase in LP horizontal momentum variance at $z=3.96$~mm [Figs.~\ref{fig:late-emitter-dynamics}(c) and~\ref{fig:late-emitter-dynamics}(d)].
An anisotropic particle distribution samples an axisymmetric field along different trajectories and can therefore develop different momentum spreads in the two transverse directions.

\section{Integrated x-ray source properties}
\label{sec:photon-properties}

\begin{figure*}[!htbp]
	\centering
	\includegraphics[width=.80\textwidth]{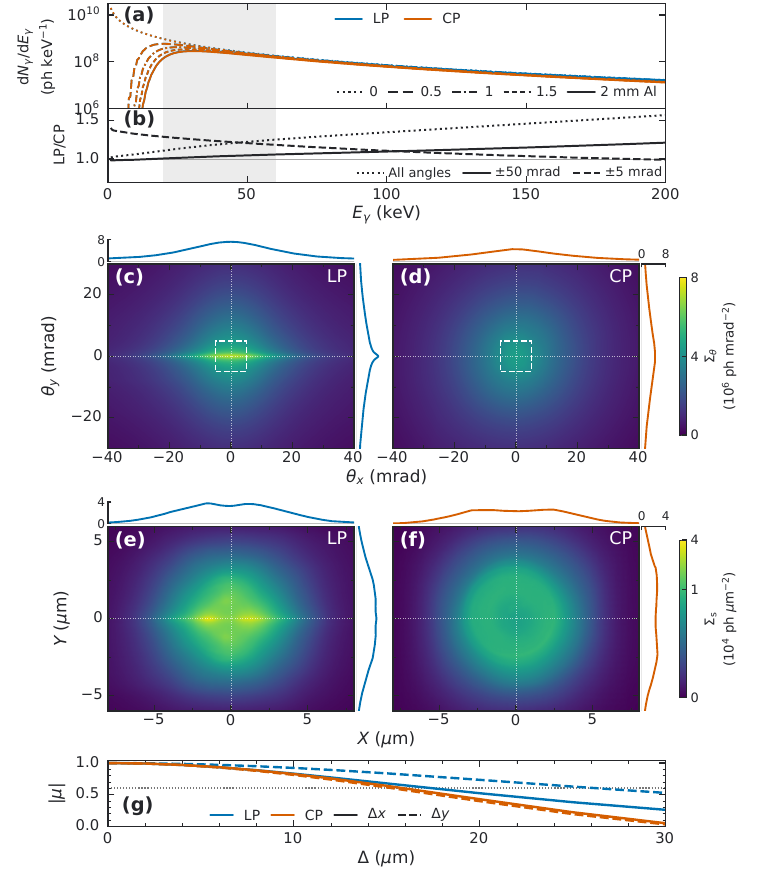}
	\caption{Betatron x rays integrated over the full interaction in the long target: LP blue/left, CP orange/right.
		(a) Photon spectra within the $\pm50$-mrad region for the indicated aluminum thicknesses, with shading marking 20--60~keV.
		(b) LP/CP spectral ratios for the indicated angular acceptances.
		(c) and (d) 20--60-keV angular fluence $\Sigma_\theta$, with dashed squares defining the central $\pm5$-mrad selection for (e) and (f).
		(e) and (f) Fluence of the effective source $\Sigma_{\mathrm{s}}$ at 35~keV with a relative bandwidth of 0.1\%, projected onto the plane $z_{\mathrm{s}}=2.7$~mm.
		Shared scales are linear in (c) and (d) and $\asinh$ in (e) and (f), and attached profiles are cuts through the origin.
		(g) Spatial coherence $|\mu|$ of 35-keV betatron x rays versus transverse separation $\Delta$, approximately 5.7~m downstream: $x$ solid, $y$ dashed, $e^{-1/2}$ dotted.
		Panels (c)--(g) use an aluminum filter of thickness 2~mm.}
	\label{fig:results_betatron}
\end{figure*}

The near equality of the central yields over $z=3.20$--$4.50$~mm does not imply equal radiation output when integrating over the full interaction.
The cumulative spectrum combines the successive emission stages, with angular acceptance selecting different proportions of each [Figs.~\ref{fig:results_betatron}(a) and~\ref{fig:results_betatron}(b)].
After transmission through 2~mm of aluminum, the integrated 20--60-keV yields within the $\pm50$-mrad square are $9.37\times10^9$ photons for LP and $9.07\times10^9$ for CP.
Restricting the acceptance to the central $\pm5$-mrad square gives $5.37\times10^8$ photons for LP and $4.36\times10^8$ for CP.
The early central LP excess described in Sec.~\ref{sec:matched-mechanism} contributes to this accumulated difference.

Increasing the aluminum thickness preferentially attenuates soft photons [Fig.~\ref{fig:results_betatron}(a)].
At each photon energy, the same transmission factor applies to LP and CP, leaving their ratio unchanged.
The ratio integrated over an energy band can nevertheless change because the filter weights the two spectra differently.

LP supplies more central 20--60-keV photons, whereas CP provides more uniform central illumination [Figs.~\ref{fig:results_betatron}(c) and~\ref{fig:results_betatron}(d)].
After transmission through 2~mm of aluminum, both beams have centered rms angular widths near 20~mrad within the $\pm50$-mrad square.
The coefficients of variation of angular fluence within the central $\pm5$-mrad square are approximately 16\% for LP and 5\% for CP.
Similar overall angular widths therefore do not imply similar illumination uniformity within the central acceptance.

The effective sources have centered rms widths of 2.3--2.9~$\mu$m and distinct directional structure in the central 35-keV channel [Figs.~\ref{fig:results_betatron}(e) and~\ref{fig:results_betatron}(f)].
A separate source calculation based on the extracted electron trajectories gives 35-keV coherence lengths of approximately 17 and 26~$\mu$m in $x$ and $y$ for LP, and 16 and 15~$\mu$m for CP [Fig.~\ref{fig:results_betatron}(g)].
For the observation geometry defined in Sec.~\ref{sec:photon-methods}, LP therefore retains coherence over larger vertical separations, while the horizontal scales are similar.

\begin{figure*}[!htbp]
	\centering
	\includegraphics[width=.84\textwidth]{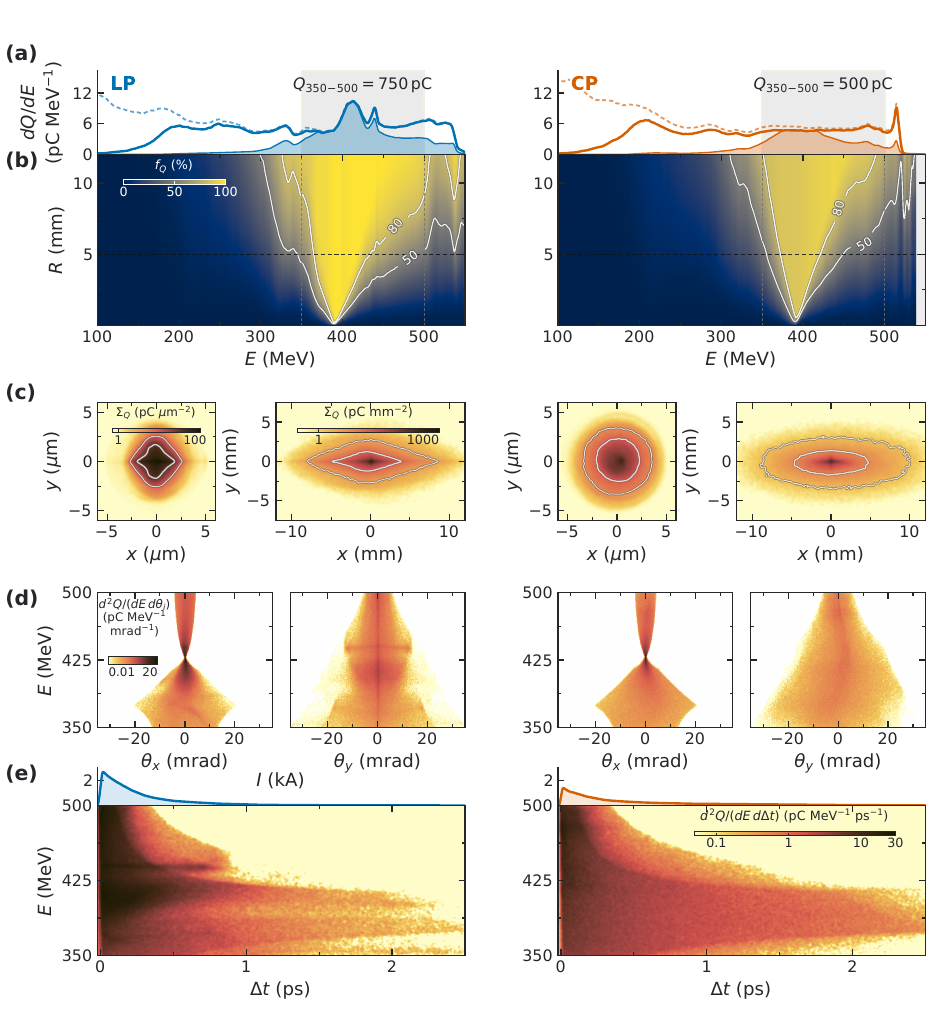}
	\caption{Transport of LP (left, blue) and CP (right, orange) electron bunches from the short target through a quadrupole triplet in a line of total length 1~m.
		(a) Spectra incident on the triplet (dashed), transmitted through it (solid), and collected at the target ($r\leq5$~mm, filled), with $Q_{350\text{--}500}$ giving collected charge selected in the shaded band of gas-exit energy.
		(b) Collection fraction $f_Q(E,R)$ versus energy and collection radius $R$, with contours marking 50\% and 80\%.
		(c) Charge fluence $\Sigma_Q$ at the gas exit (left subpanels) and target (right subpanels), with contours enclosing 50\% (inner) and 80\% (outer) of the charge.
		(d) Target energy--angle densities.
		(e) Target current (upper) and joint density in energy and arrival time (lower).
		In (a) and (b), $E$ is the electron kinetic energy at the gas exit, before transport.
		Panels (c)--(e) select $350\leq E<500$~MeV at the plane shown, with $r\leq5$~mm in (d) and (e).}
	\label{fig:vhee-transport}
\end{figure*}

\section{Electron extraction and transport}
\label{VHEE}

The short target provides a complementary test of how the extracted electron phase-space distribution affects the charge delivered after transport.
Its bunches leave the plasma before the interval used to select the radiating electrons in Sec.~\ref{sec:angular-acceptance}.
We reconstruct each forward electron's crossing of the gas-exit plane while preserving correlations among position, momentum, and crossing time.
We use ImpactX~\cite{Huebl2022} to track the full bunch ($E\gtrsim50$~MeV) through an aligned, idealized transport line of length 1~m containing a quadrupole triplet.
The model includes apertures, nonlinear fringe fields, and space charge.
Both polarizations use the same magnet settings.
Energy and radius cuts are applied at the target after transport.
The populated 350--500-MeV band at the gas exit and $R=5$~mm collection radius define the delivery figure of merit, with the radius dependence also retained in Fig.~\ref{fig:vhee-transport}(b).
We selected the triplet geometry and gradients by Bayesian optimization with an ensemble surrogate and an expected-improvement criterion~\cite{Jones1998}, maximizing the minimum LP/CP collection fraction within 5~mm under transmission and pointing constraints.
Appendices~\ref{app:transport} and~\ref{app:impactx-parameters} give details on the full numerical model.

The delivered charges reported in the text and tables are selected by energy at the target, and $\Sigma_Q=\mathrm{d}^2Q/(\mathrm{d}x\,\mathrm{d}y)$ denotes transverse charge fluence.
Figures~\ref{fig:vhee-transport}(a) and~\ref{fig:vhee-transport}(b) instead use energy at the gas exit, including the charge labels in panel (a).
Here $f_Q(E,R)$ is the fraction of incident charge collected within target radius $R$ in each interval of incident energy.

LP delivers more charge through the triplet despite nearly equal total extracted charge above 100~MeV.
The gas-exit charges above this threshold are 2957~pC for LP and 2949~pC for CP (Appendix~\ref{app:electron-evolution}), but their incident 350--500-MeV band charges are 958 and 797~pC, respectively (Table~\ref{tab:vhee-peak}).
At the target, the $350\leq E<500$~MeV and $r\leq5$~mm selection contains 750~pC for LP and 500~pC for CP [Figs.~\ref{fig:vhee-transport}(a)--\ref{fig:vhee-transport}(c)], corresponding to approximately 78\% and 63\% of the respective incident band charges.
In the modeled line, LP therefore combines a larger incident band population with more effective collection of its transverse phase space.

The two bunches have similar mean energies near 425~MeV, but different sizes, divergences, and correlations at the gas exit (Table~\ref{tab:vhee-peak}).
Their target transverse sizes remain comparable on the millimeter scale.
Chromatic focusing depends on momentum, giving electrons of different energies different trajectories [Figs.~\ref{fig:vhee-transport}(b) and~\ref{fig:vhee-transport}(d)].
A narrow angular distribution can consequently coexist with a broad spot.
Appendix~\ref{app:transport} relates these correlations to projected emittance growth.

The CP bunch is shorter than the LP bunch at the gas exit but longer after collection.
The rms duration of the initially shorter CP bunch increases to 0.54~ps at the target, compared with 0.33~ps for LP [Fig.~\ref{fig:vhee-transport}(e)].
This stretching is consistent with different path lengths, while speed differences along straight paths contribute only a spread on the femtosecond time scale.

\section{Discussion and conclusions}
\label{sec:discussion}
\label{conclusions}

CP produces more photons early in the long target, whereas LP produces more later.
The factorization of critical energy associates the accompanying spectral hardening with the joint evolution of electron energy and transverse bending, rather than electron energy alone.
For the fixed cohort selected by its contribution to the later emission, the LP transverse momentum distribution develops a larger horizontal than vertical spread.
At the peak of its horizontal momentum spread, the variance increase is dominated by momentum changes produced by axisymmetric fields along the actual trajectories, connecting the field decomposition to the observed motion.
Earlier work connects driver polarization to injected phase space~\cite{Schroeder2014,Tomassini2022} and to the polarization of betatron x rays~\cite{Vieira2016}.
The present comparison follows how those electron dynamics affect photon production and angular collection during propagation.
Stronger and harder emission can be accompanied by angular broadening, so the polarization that produces more photons does not necessarily deliver more within a fixed aperture.

Experimental implementation requires assessments of stability between shots and alignment tolerances, modeling of the laser dump material, and precise magnetic field mapping.
An imaging model could propagate the calculated x-ray distributions through the filter, interferometer, object, and detector to evaluate mammographic image quality and dose.
The transported electron distributions could provide inputs to dose calculations for a specified VHEE delivery geometry and target.

Within the configurations studied, the relative benefits of LP and CP depend on the propagation stage and collection acceptance.
CP produces stronger early betatron emission, whereas LP produces stronger emission later, but horizontal angular broadening leaves their central filtered 20--60-keV yields nearly equal during this later stage.
Integrated over the full interaction, LP supplies more central photons while CP provides more uniform illumination.
The short target gives almost equal total charge above 100~MeV for the two polarizations, yet LP delivers more charge in the 350--500-MeV band through a quadrupole triplet.

\section*{Acknowledgments}
The authors thank the ELI-NP information technology (IT) department for its continued support of the ELI-NP graphics processing unit (GPU) cluster, and acknowledge useful discussions with M. Ciubancan and D. Zabet.
A.C.B. and V.H. acknowledge the EuroHPC Joint Undertaking for awarding this project access to the EuroHPC supercomputer LUMI, hosted by CSC (Finland) and the LUMI consortium through a EuroHPC Regular Access call under project EHPC-REG-2026R01-063 (\texttt{\detokenize{project_465003258}}).

V.H., D.S., and P.T. acknowledge support from the Romanian Government and the Health Program, within the project “Medical applications of high-power lasers - Dr. LASER” (SMIS code 326475).
A.C.B. acknowledges support from the Extreme Light Infrastructure Nuclear Physics Phase II project, co-financed by the Romanian Government and the European Union through the European Regional Development Fund and the Competitiveness Operational Programme (No.~1/07.07.2016, COP, ID~1334) and ELI-RO/DEZ/2023\_001, funded by the Romanian Ministry of Education and Research.
P.T. acknowledges additional support from ELI-RO/14 SPARC, funded by the Romanian Ministry of Education and Research.

\par\medskip\noindent
A.C.B. developed the computational methodology and diagnostic software, including radiation diagnostics for simulations in a boosted frame, performed the production particle-in-cell simulations and beam transport calculations, conducted numerical validation, analyzed and interpreted the results, prepared the figures, and led the writing of the manuscript.
V.H. and D.S. contributed to formal analysis and manuscript preparation.
P.T. designed the initial accelerator configurations and performed the associated particle-in-cell simulations.
All authors reviewed and approved the final manuscript.

\par\medskip\noindent
The authors declare no competing interests.

\begin{samepage}
\section*{Data availability}
Simulation inputs, analysis and plotting scripts, and full raw simulation outputs are available upon reasonable request. The accompanying Zenodo repository~\cite{Berceanu2026Data} is not yet publicly accessible.
\par
\end{samepage}

\appendix
\renewcommand{\appendixname}{APPENDIX}
\section{ELECTRON EXTRACTION AND TRANSPORT DIAGNOSTICS}
\label{app:transport}
\label{sec:beamline}

We model the electron branch shown in Fig.~\ref{fig:setup}(b) up to a target located 1~m downstream of the gas exit and upstream of the dipole spectrometer.
The three permanent-magnet quadrupoles have lengths of 127.6, 141.6, and 89.9~mm and gradients of $+108.3$, $-105.6$, and $+101.1$~T\,m$^{-1}$, respectively.
Each has a clear radius of 14.3~mm.
The entrance drift, two gaps between magnets, and final drift are 72.8, 21.2, 31.1, and 515.8~mm, respectively.
For a quadrupole gradient $g_{\rm q}$, the fields are $B_x=g_{\rm q}y$ and $B_y=g_{\rm q}x$.
With this sign convention, the sequence is defocusing--focusing--defocusing in $x$ and focusing--defocusing--focusing in $y$.
The model includes nonlinear fringes, apertures, and space charge (Appendix~\ref{app:impactx-parameters}).

Crossings at the gas exit are reconstructed by matching species and particle IDs between laboratory snapshots, using cubic Hermite position interpolation and linear momentum interpolation.
Table~\ref{tab:vhee-peak} describes the complete 350--500-MeV source band.
The beam-optics slopes are $i'=p_i/p_z$, with $i=x,y$.
Primes here denote slopes, not quantities in the Lorentz-boosted frame.
Original weights and position--momentum--time correlations are retained.
The transport calculation starts from these distributions at the gas exit and omits subsequent interactions with residual laser and plasma fields and with material in the laser dump.

For Fig.~\ref{fig:vhee-transport}(e), the arrival coordinate is $\Delta t=t_{\rm target}-\bar t_{\rm source}-L/c$, with line length $L$ and mean crossing time $\bar t_{\rm source}$ for the incident band, weighted by charge.
Integrating the joint density in energy and arrival time over the selected band gives the current $I(\Delta t)=\mathrm{d}Q/\mathrm{d}t_{\rm target}$.
Target durations and currents are calculated for electrons with $350\leq E<500$~MeV and $r\leq5$~mm at the target, whereas the source duration in Table~\ref{tab:vhee-peak} includes the full incident band.

\begin{table}[!b]
	\caption{Electron properties at the short target's gas exit ($z=2.46$~mm) for forward crossings with $350\leq E<500$~MeV, integrated over all angles and transverse positions.
	Moments are weighted by charge, rms widths are centered, divergence uses slopes $i'=p_i/p_z$, emittances are projected, and $\sigma_t$ measures the spread in crossing time.}
	\label{tab:vhee-peak}
	\centering
	\small
	\setlength{\tabcolsep}{3pt}
	\begin{tabular}{p{.53\columnwidth}rr}
		\toprule
		Quantity & LP & CP \\
		\midrule
		Charge $Q_{350\text{--}500}$ (pC) & 958 & 797 \\
		Mean kinetic energy $\langle E\rangle$ (MeV) & 426.5 & 424.4 \\
		Rms energy spread $\sigma_E$ (MeV) & 40.5 & 43.7 \\
		Relative rms spread $\sigma_E/\langle E\rangle$ (\%) & 9.49 & 10.28 \\
		Rms size $\sigma_x,\sigma_y$ ($\mu$m) & 1.29, 1.38 & 1.84, 1.81 \\
		Rms divergence $\sigma_{x'},\sigma_{y'}$ (mrad) & 12.85, 13.44 & 17.99, 17.87 \\
		\raggedright Normalized rms emittance $\epsilon_{n,x},\epsilon_{n,y}$ (mm mrad) & 11.15, 12.75 & 21.30, 20.98 \\
		\raggedright Rms spread in crossing time $\sigma_t$ (fs) & 7.83 & 5.73 \\
		\bottomrule
	\end{tabular}
\end{table}

\begin{figure}[!htbp]
\centering
\includegraphics[width=\columnwidth]{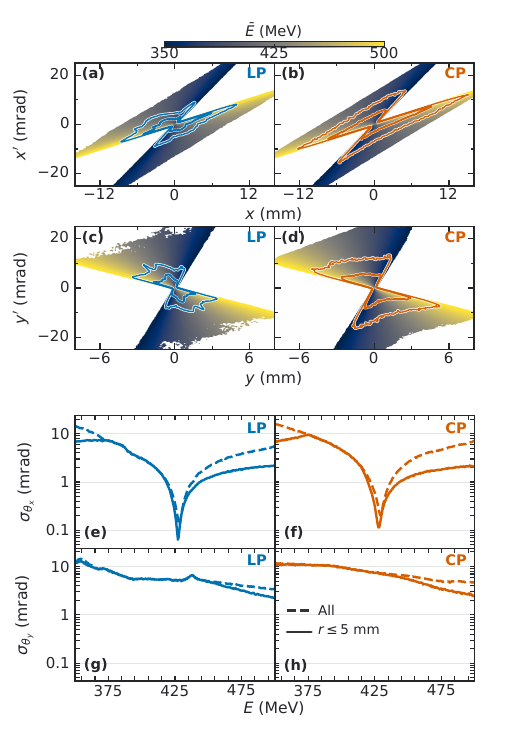}
\caption{Transverse electron distributions at the target 1~m downstream of the gas exit, for $350\leq E<500$~MeV.
(a)--(d) Phase spaces of all transmitted electrons, horizontal in (a) and (b) and vertical in (c) and (d), with slopes $i'=p_i/p_z$ ($i=x,y$).
Color shows mean energy $\bar{E}$ weighted by charge, and contours enclose 50\% (thick) and 80\% (thin) of the charge.
(e)--(h) Corresponding centered rms angular widths, weighted by charge, with $\theta_i=\tan^{-1}i'$.
Dashed curves include all transmitted electrons, while solid curves select $r\leq5$~mm.}
\label{fig:transport-phase-space}
\label{fig:transport-divergence}
\end{figure}

Figure~\ref{fig:transport-phase-space} resolves the transverse phase space of the delivered 350--500-MeV electrons.
The horizontal rms divergence has a pronounced minimum near 430~MeV, while the vertical divergence remains several milliradians at that energy.
The position--angle correlations show why a narrow angular distribution can retain a finite spot size.

\phantomsection\label{app:transport-emittance-interpretation}
For each selection in Table~\ref{tab:transport-emittance}, we calculate the source and target emittances from the same particles using
\begin{equation}
\epsilon_{n,i}^2=\langle(\delta i)^2\rangle\langle(\delta u_i)^2\rangle-\langle\delta i\,\delta u_i\rangle^2,
\end{equation}
where $i=x,y$, $u_i=p_i/(m_ec)$, and $\delta$ denotes deviations from means weighted by charge.
These cohorts are selected at the target and differ from the incident band in Table~\ref{tab:vhee-peak}.

\begin{table*}[!htbp]
\caption{Transmitted charge and projected normalized rms emittances at the gas exit and the target 1~m downstream, with units given in the column headings.
Rows comprise all transmitted electrons with $350\leq E<500$~MeV at the target or the subset with $r\leq5$~mm.}
\label{tab:transport-emittance}
\centering
\small
\setlength{\tabcolsep}{4pt}
\begin{tabular*}{\linewidth}{@{\extracolsep{\fill}}llrrrrr@{}}
\toprule
 & Target & $Q_{350\text{--}500}$ & \multicolumn{2}{c}{Source} & \multicolumn{2}{c}{Target} \\
Case & selection & (pC) & $\epsilon_{n,x}$ & $\epsilon_{n,y}$ & $\epsilon_{n,x}$ & $\epsilon_{n,y}$ \\
 & & & \multicolumn{2}{c}{(mm~mrad)} & \multicolumn{2}{c}{($10^3$~mm~mrad)} \\
\midrule
LP & all & 934 & 10.1 & 12.6 & 10.3 & 6.32 \\
LP & $r\leq5$~mm & 750 & 7.56 & 12.1 & 5.01 & 4.79 \\
CP & all & 719 & 17.8 & 21.4 & 17.4 & 11.0 \\
CP & $r\leq5$~mm & 500 & 13.6 & 21.6 & 7.34 & 7.85 \\
\bottomrule
\end{tabular*}
\end{table*}

We tested the origin of the projected emittance growth by applying paraxial linear transport matrices evaluated at each electron's energy to the same cohorts selected at the target.
Without nonlinear corrections or space charge, this control reproduces all target emittances in Table~\ref{tab:transport-emittance} to within 2.4\%.
Applying a single transfer matrix evaluated at 425~MeV to every electron instead preserves the source emittances.
The large projected growth therefore predominantly reflects chromatic focusing, which superposes differently correlated energy slices.

\section{NUMERICAL IMPLEMENTATION AND CONVERGENCE TESTS}
\label{app:numerics}

\begin{figure}[!htbp]
  \centering
  \includegraphics[width=\columnwidth]{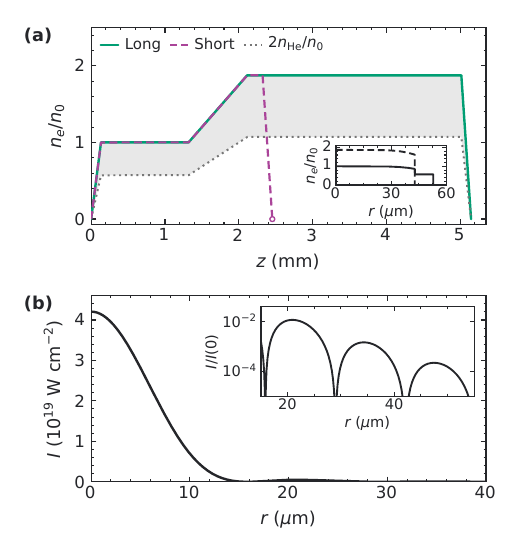}
  \caption[Target and laser profiles.]{Target and laser profiles, common to LP and CP.
(a) Initial free-electron density $n_e/n_0$ on axis, with reference $n_0=6.72\times10^{18}$~cm$^{-3}$ in the first plateau.
The dotted curve and gray shading decompose the profile of the long target into $2n_{\rm He}/n_0$ and $5n_{\rm N}/n_0$, with both species extending to the gas exit.
The open circle marks the exit of the short target.
Inset: radial profiles at $z=1.0$~mm (solid) and $2.2$~mm (dashed).
(b) Laser intensity averaged over the optical cycle at the vacuum focus and pulse maximum.
Inset: $I/I(0)$ on a logarithmic scale.}
  \label{fig:gas-density}
\end{figure}

\textit{Target.}\phantomsection\label{app:fbpic-parameters}
The targets differ only in the length of the second plateau (Fig.~\ref{fig:gas-density}).
Helium and nitrogen are initialized as He$^{2+}$ and N$^{5+}$, with a ratio of atomic number densities $n_{\rm N}/n_{\rm He}=0.3$ throughout the gas core.
Nitrogen ionization follows the Ammosov--Delone--Krainov (ADK) model~\cite{ADK1986}.
The initial free-electron density is $n_e=2n_{\rm He}+5n_{\rm N}$, with axial value $n_0=6.72\times10^{18}$~cm$^{-3}$ in the first plateau.
The on-axis total rises linearly from $n_0$ to $1.26\times10^{19}$~cm$^{-3}$ between $z=1.32$ and $2.11$~mm.
Both species retain the same longitudinal shape through the exit ramp.
Both targets have linear entrance and exit ramps, each 0.13~mm long, with total target lengths of 2.46~mm for the short target and 5.15~mm for the long target.
The radial density factor is $\exp\{-2[r/(66\,\mu\mathrm{m})]^6\}$ for $r<42.8~\mu$m, with an untapered helium annulus from $42.8$ to $52.8~\mu$m for $z\leq1.85$~mm.

\smallskip\textit{Laser.}
The pulses share nominal energy $2.1$~J, wavelength $\lambda_0=810$~nm, and vacuum focus $z_f=0.53$~mm.
The assumed target energy of 2.1~J accounts for transmission losses from a compressed pulse of 2.5~J.
FBPIC's flattened Gaussian profile~\cite{Santarsiero1997} uses order $N=16$ and transverse scale $w_0=8.25~\mu$m, giving an intensity FWHM diameter of $13~\mu$m at focus.
The order $N$ controls the profile flatness far from focus.
The parameter $\tau=21.23$~fs of the Gaussian temporal envelope corresponds to an intensity FWHM duration $\sqrt{2\ln2}\,\tau=25$~fs.
LP uses one component with normalized amplitude $a_0=4.49$, while CP uses two orthogonal components with $a_0=3.18$ and relative phase $\pi/2$.
Both have the same peak intensity of $4.2\times10^{19}$~W~cm$^{-2}$ when averaged over the optical cycle.

\smallskip\textit{FBPIC numerics.}\phantomsection\label{app:fbpic-grid}
For frame transformations, primes denote the boosted simulation frame and unprimed quantities denote the laboratory frame.
The quasi-cylindrical FBPIC~0.27.0 simulations~\cite{Lehe2016} use a Lorentz-boosted frame ($\gamma_b=2$), a 32nd-order Galilean pseudospectral solver ($v'_{\rm Gal}=-\beta_b c$), the Vay pusher, cubic particle shapes, and single-pass compensated binomial filtering.
The grid in the boosted frame has $3365\times1792$ cells in $(z',r)$ over a domain $291~\mu$m long and $182~\mu$m in radius, excluding guard and damping cells, with four azimuthal modes ($m=0,\ldots,3$) and a time step of $274$~as.
Boundaries are longitudinally open and radially reflective.
The laboratory window velocity is $v_w\simeq0.999\,c$.
All plotted propagation histories use the coordinate of the window front, $z=v_wt$, initially zero, approximated by $ct$ in the main text.
Each species initially has $2\times2\times16$ particles in $(z',r,\theta)$ per core cell and $1\times1\times16$ in the outer annulus.
Laboratory fields and charge density are saved every $92$~fs, and electron spectra every $23$~fs for $u_z>100$.
Additional diagnostics track sampled trajectories and measure current and momentum flux moments over the full population.

\smallskip\textit{Radiation model.}\phantomsection\label{app:radiation-model}
The newly developed radiation diagnostic in the boosted frame extends FBPIC's synchrotron module for the laboratory frame~\cite{FBPICbetatron}, retaining incoherent emission in the strong-wiggler regime.
At each PIC step, we transform electron momenta and local fields to the laboratory frame and integrate the emission over $\mathrm{d}t=(\gamma/\gamma')\,\mathrm{d}t'$, where $\gamma$ and $\gamma'$ are the particle Lorentz factors in the laboratory and boosted frames, respectively.
Events enter propagation bins at $t=\gamma_b(t'+\beta_b z'/c)$, coupling boosted time and position.
For each species, the diagnostics provide spectra as a function of time, maps of emission and effective sources, and joint position--angle moments.
Thresholds are $\gamma>100$ and a minimum critical energy of 2~keV.
The photon grid spans 1--200~keV at 0.5-keV spacing and $\pm50$~mrad per coordinate, with angular spacing of 0.25~mrad for Figs.~\ref{fig:matched-mechanism} and~\ref{fig:results_betatron}, and 0.5~mrad for convergence tests.
Radiation reaction is disabled.

\smallskip\textit{Sampling and coherence.}
Photon spectra use bins of 25~fs in laboratory time, whereas angular and source maps use bins of 125~fs.
Source pixels are 0.125~$\mu$m wide.
For coherence in Eq.~(\ref*{eq:transverse-coherence}), trajectories are sampled with probability $1/512$ for electrons released by nitrogen ionization and $1/4096$ for the other electron populations, retaining statistical weights.
States are saved every PIC step for laboratory $\gamma>100$ and at least every fourth step otherwise, including birth states and their laboratory times.
We integrate the classical radiation amplitude using a piecewise-linear radiation factor in retarded time, retaining spherical propagation phases and evaluating propagation directions and distances at segment midpoints.
\par

\smallskip\textit{Aluminum attenuation.}\phantomsection\label{app:aluminum-filter}
Aluminum thicknesses $d=0$, $0.5$, $1$, $1.5$, and $2$~mm are evaluated by multiplying photon weights by the Beer--Lambert transmission $T_d(E_\gamma)=[T_{2\,\mathrm{mm}}(E_\gamma)]^{d/(2\,\mathrm{mm})}$ for $d>0$, with $T_0=1$ for the unfiltered case.
The tabulated transmission through 2~mm of aluminum is linearly interpolated in energy and extended above 150~keV using mass attenuation coefficients from the National Institute of Standards and Technology (NIST)~\cite{HubbellSeltzer2004}, interpolated in log--log space, using an aluminum density of $2.7$~g\,cm$^{-3}$.
The model describes primary transmission through a uniform slab at normal incidence, preserving transmitted photon energies and directions and treating absorption and scattering as losses.

\begin{figure*}[!htbp]
\centering
\includegraphics[width=.78\textwidth]{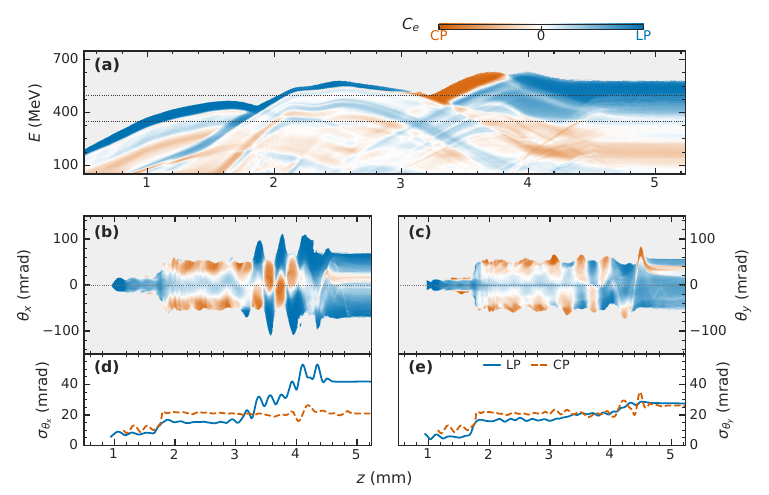}
\caption{Spectral and angular evolution of LP and CP electron distributions in the long target, with $u_z>100$.
(a) Contrast $C_e=(D^{\rm LP}-D^{\rm CP})/(D^{\rm LP}+D^{\rm CP})$ of the charge spectra, $D=\mathrm{d}Q/\mathrm{d}E$, with dotted lines delimiting $350\leq E<500$~MeV.
(b) and (c) Angular contrasts in this band, selected independently at each snapshot, with $D=\mathrm{d}Q/\mathrm{d}\theta_x$ and $\mathrm{d}Q/\mathrm{d}\theta_y$, respectively.
Blue (orange) denotes LP (CP) excess.
(d) and (e) Centered rms angular widths $\sigma_{\theta_x}$ and $\sigma_{\theta_y}$ weighted by charge, including all angular tails, for LP (blue solid) and CP (orange dashed).}
\label{fig:electron-evolution}
\end{figure*}

\begin{table*}[!htbp]
	\caption{Electron and photon properties versus ellipticity $\chi$.
		Electron quantities use $E>100$~MeV at the final snapshot for the long target (within the saved $u_z>100$ population) or at the gas exit for the short target.
		Within the $\pm50$-mrad square, $N_{20\text{--}60}$ counts accumulated 20--60-keV photons, and Al denotes transmission through 2~mm of aluminum.
		The median energy $E_{\gamma,50}$ is the photon energy below which 50\% of the unfiltered radiated energy in the 1--200-keV band is emitted.
		Dashes indicate that radiation diagnostics were disabled.}
	\label{tab:polarisation}
	\centering
	\small
	\setlength{\tabcolsep}{4pt}
	\begin{tabular*}{\linewidth}{@{\extracolsep{\fill}}llccccc@{}}
		\toprule
		Profile & $\chi$ & $Q_{>100}$ & $\sigma_{\theta_x},\sigma_{\theta_y}$ & $N_{20\text{--}60}$ & $N_{20\text{--}60}^{\rm Al}$ & $E_{\gamma,50}$ \\
		 & & (pC) & (mrad) & ($10^9$ ph) & ($10^9$ ph) & (keV) \\
		\midrule
		Long & 0 (LP) & 1935 & 55.00, 34.09 & 16.56 & 9.37 & 54.5 \\
		Long & 0.5 & 1912 & 41.23, 35.30 & 14.96 & 8.46 & 53.2 \\
		Long & 1 (CP) & 2184 & 32.53, 36.62 & 16.11 & 9.07 & 51.0 \\
		Short & 0 (LP) & 2957 & 21.01, 22.28 & \textemdash & \textemdash & \textemdash \\
		Short & 0.5 & 2942 & 24.24, 23.22 & \textemdash & \textemdash & \textemdash \\
		Short & 1 (CP) & 2949 & 25.84, 25.95 & \textemdash & \textemdash & \textemdash \\
		\bottomrule
	\end{tabular*}
\end{table*}

\smallskip\textit{Convergence tests.}\phantomsection\label{app:fbpic-convergence}
Tests of grid and time step at fixed domain size cover both targets and polarizations using a coarser $2692\times1344$ grid and a step of $342$~as.
Tests for the long target on the baseline grid use five azimuthal modes or increased loading of $3\times3\times16$ (core) and $2\times2\times16$ (outer annulus) ppc (particles per cell).
For the short target, electron spectra and the evolution of charge and energy change by less than $5\%$.
All three refinements change the LP and CP photon spectra and filtered 20--100-keV fluence (2~mm Al) for the long target by less than $5\%$.
Differences are measured using $L_1$ norms for spectra and fluence and $L_2$ norms for charge and energy evolution.

\smallskip\textit{ImpactX transport.}\phantomsection\label{app:impactx-parameters}
ImpactX~26.08~\cite{Huebl2022} imposes nominal on-axis injection by subtracting the transverse centroid and mean slopes of the 350--500-MeV population, weighted by charge, from the full bunch at the gas exit ($E\gtrsim50$~MeV), preserving particle energies and weights.
Fourth-order symplectic integration advances the exact Hamiltonian in the quadrupole bodies, with nonlinear thin fringe maps~\cite{Forest1988}.
Slices are $\leq0.4$~mm in quadrupoles (two steps each) and $\leq2.5$~mm in drifts.
Space charge uses a solver based on an integrated Green function with open boundaries, fast Fourier transforms, a dynamic $96^3$ mesh, and cubic particle shapes.

\smallskip\textit{Hardware.}
Each FBPIC run for the long target takes approximately 7~h on 20 NVIDIA H100 GPUs with 80~GB each (five nodes, one Message Passing Interface (MPI) rank per GPU).
The Bayesian optimization of the quadrupole triplet and all final ImpactX transport calculations were performed on LUMI using central processing units (CPUs).
\section{ELECTRON ANGULAR EVOLUTION AND LASER-ELLIPTICITY DEPENDENCE}
\label{app:electron-evolution}
\label{sec:ellipticity}

Figure~\ref{fig:electron-evolution} relates the LP--CP differences in the spectra to transverse angular broadening in the simulations of the long target in Fig.~\ref{fig:Betatron_global}.
Distributions are compared at matched laboratory times using the normalized contrast
\begin{equation}
C_e[D]=\frac{D^{\rm LP}-D^{\rm CP}}{D^{\rm LP}+D^{\rm CP}},
\end{equation}
where $D=\mathrm{d}Q/\mathrm{d}E$ for spectra and $D=\mathrm{d}Q/\mathrm{d}\theta_i$ for angular distributions, with $\theta_i=\tan^{-1}(p_i/p_z)$, $i=x,y$.

The angular distributions include electrons with $350\leq E<500$~MeV at each snapshot, so the population changes as electron energies evolve.

Near the position where the short target ends, $z=2.46$~mm, the LP and CP distributions in the long target have horizontal rms widths of 16 and 21~mrad, respectively.
By $z=5.24$~mm, the LP width has increased to 42~mrad while the CP width remains 21~mrad.
At $z=5.24$~mm, their vertical rms widths are 28 and 26~mrad, respectively.
The larger final LP band charge therefore accompanies predominantly horizontal broadening, consistent with the motion of the late emitters.

Table~\ref{tab:polarisation} extends the LP--CP comparison to intermediate ellipticity $\chi=0.5$, at the same nominal pulse energy and separately for each target length.
The charges from the short target differ by less than 1\% across the three cases.
At $\chi=0.5$, the 20--60-keV photon yield from the long target is approximately 7--10\% lower than the yields at $\chi=0$ and $\chi=1$.
Thus similar charge from the short target can coexist with a nonmonotonic radiation response across the chosen ellipticities.

\pdfbookmark[1]{References}{references}
\makeatletter
\let\auto@bib@innerbib\@empty
\makeatother

\end{document}